%% file: wire4.tex
\documentclass[final]{IEEEtran}
\usepackage{graphicx}
\usepackage[namelimits,intlimits,sumlimits]{amsmath}
\usepackage{amsfonts}
\usepackage{txfonts}
\usepackage{amssymb}
\usepackage{psfrag}

\begin{document}

\newcommand{\ud}{d}

\title{Analytical model of nanowire FETs in a partially ballistic or
  dissipative transport regime}

\author{Paolo Michetti, Giorgio Mugnaini, Giuseppe Iannaccone~\IEEEmembership{Member,~IEEE,}%
  \thanks{This work was supported in part by the EC Nanosil FP7 Network of 
    Excellence under Contract 216171 and in part by the European Science 
    Foundation EUROCORES Programme Fundamentals of Nanoelectronics, through 
    fundings from Consiglio Nazionale delle Ricerche (awarded to 
    IEEIIT-PISA) and the European Commission Sixth Framework Programme under 
    Project Dewint (Contract ERAS-CT-2003-980409).}%
  \thanks{Authors are with the Dipartimento di Ingegneria dell'Informazione,
    Universit\'a di Pisa, Pisa I-56122, Italy (e-mail:
    p.michetti@iet.unipi.it; g.inannaccone@iet.unipi.it).}
}

\maketitle

\begin{abstract}
  The intermediate transport regime in nanoscale transistors between
  the fully ballistic case and the quasi-equilibrium case described by
  the drift-diffusion model is still an open modeling issue.
  Analytical approaches to the problem have been proposed, based on the
  introduction of a backscattering coefficient, or numerical approaches
  consisting in the Monte Carlo solution of the Boltzmann transport
  equation or in the introduction of dissipation in quantum transport
  descriptions.
  In this paper we propose a very simple analytical model to seamlessly
  cover the whole range of transport regimes in generic quasi-one 
  dimensional
  field-effect transistors, and apply it to silicon nanowire transistors.
  The model is based on describing a generic transistor as a chain of 
  ballistic
  nanowire transistors in series, or as the series of a ballistic 
  transistor
  and a drift-diffusion transistor operating in the triode region.
  As an additional result, we find a relation between the
  mobility and the mean free path, that has deep consequences
  on the understanding of transport in nanoscale devices.
\end{abstract}


\markboth{} {Mugnaini, Michetti and Iannaccone: Ballistic chain description of the transport in SNWTs}

\section{INTRODUCTION}

Multiple gate architectures such as gate-all-around (GAA) MOSFETs have lately
attracted significant interest \cite{lauhon,singh,suk}, and have emerged 
as promising options to keep short channel effects under control, exhibiting
quasi-ideal subthreshold swing with undoped channels.
This has the very important consequence of alleviating intrinsic variability of
transistor threshold voltage, which in planar MOSFETs is mainly due to channel doping. 

Nanowire FETs are a particular case of multiple gate FETs, in which 
quantum confinement occurs in the transverse cross section of only few nanometers. Nanowire FETs are basically
quasi one-dimensional transistors, where transport occurs in a set of loosely
coupled propagating modes.

From the point of view of modeling, several papers have appeared in the literature 
addressing transport and quantum confinement in silicon nanowire
transistors.
In pioneering works \cite{sanders93,shen93,baie,colinge}, the effects of quantum confinement on
a silicon nanowire were discussed. 
Numerical detailed investigations of quantum confinement
in silicon and silicon germanium nanowires with the anisotropic effective mass approximation, and its effect in lifting the degeneracy of silicon
conduction band minima were discussed in \cite{curatola2002,trellakis}.
The electrostatics of silicon nanowire devices with cylindrical symmetry has been investigated
through a perturbative approach to the Schr\"odinger equation \cite{gnani} or with a self-consistent
solution in Poisson-Schr\"odinger equation with cylindrical coordinates
\cite{pokatilov}.

Analytical models of ballistic nanowire transistors have been proposed
in \cite{jimenez03,jimenez04,iniguez05}
and a broad review can be found in \cite{iniguez2006}.
In real nanowire devices currents are much lower than those
predicted by ballistic models \cite{singh}, which can only be used as an
asymptotic performance limit.

Non ballistic transport in quasi-1D 
channels is harder to model. 
As far as numerical studies
are concerned, far-from-equilibrium transport in silicon nanowire
transistors was investigated in \cite{wang} within the non-equilibrium
Green's functions formalism, for both ballistic and dissipative transport, 
using the B\"uttiker probes approach to model inelastic scattering.
A subband-based drift-diffusion simulation, in which the 3D electrostatics
is solved self-consistently with the 2D Schr\"odinger equation in each transverse
cross section and a set of 1D continuity equations based on the drift-diffusion 
description, has been proposed in \cite{fiori2007}.

As far as analytical models of dissipative transport in quasi-1D FETs are concerned, 
notable examples are Ref. \cite{jimenez04}, which proposed a semiclassical model
with drift-diffusion transport and constant mobility inside a cylindrical
MOSFET, and Ref. \cite{paul}, in which a polynomial expansion
of the Fermi integrals for the mobile charge is used.

Specific scattering mechanisms such as phonon scattering
have been numerically addressed within the non-equilibrium Green's 
functions approach by Jin {\em et al.} \cite{Jin} and by M.~Gilbert
{\em et al.} \cite{gilbert,gilbert2}.

We believe it would be very interesting to have an analytical model
capable to seamlessly cover the continuum of transport regimes between
the limits of ballistic transport and drift-diffusion (i.e. quasi-equilibrium)
transport.
Such a model, theoretically derived from the formalism of B\"uttiker
virtual probes \cite{buttiker},
and consisting in either a chain of ballistic transistors or in the series of a fully ballistic 
and an ideal drift-diffusion transistor, has been proposed in \cite{mugnaini,mugnaini2}
for 2D MOSFETs.

On the basis of this work, in the present paper we present an analytical model capable of 
describing the complete range of transport regimes in quasi-1D FETs, from fully ballistic to long 
channel quasi-equilibrium drift-diffusion behavior. 
A preliminary attempt has been proposed in \cite{mugnaini_ESSDERC}. 
As we shall show, the model is sufficiently simple to be suitable for circuit-level simulations 
and provides a strong intuitive picture of the transition from ballistic to drift-diffusion transport, 
which is missing in other descriptions of partially ballistic transport such as those relying
on the introduction of a backscattering coefficient \cite{lundstrom97,rahman}.

The paper is organized as follows: in Section II we set up a
model for ballistic transport in a nanowire transistor, that, in Section
III, we apply to the case of a chain of ballistic transistors.
With a linearization procedure we show that a sufficiently long chain of
ballistic transistors, can be regarded as a drift-diffusion channel.
But the same approach fails for short ballistic
chains, in which transport has an intermediate nature between
ballistic and drift-diffusion.
This difficulty is tackled in Section IV, where we present a
compact model for intermediate transport that treats the ballistic chain
as a series of one drift-diffusion section, for the first $N-1$
transistors, and the remaining one ballistic channel, in which the
non-equilibrium character of the intermediate transport manifests itself.
In Section V we introduce the cylindrical and rectangular
confinements for silicon nanowires considered in this paper and in Section VI we compare the results of our drift-diffusion
and intermediate transport compact models with the numerical solution of the transport through ballistic chains of different length.
In the section we also give an estimation of the current ballisticity ratio as a function of
the transistor chain length.

\section{BALLISTIC TRANSPORT}
\input{figurea.tex}

In the following discussion, we describe our approach in the general
situation of a n-FET with a quasi-1D channel, with the effective mass approximation.
Indeed the subband energies are determined by the transverse confinement, and to explicitly account for the capacitive coupling between gate and
channel, the contact geometry has to be taken in account.   
The bottom of the 1D conduction subbands are formally defined as the
sum $q \varepsilon_{\alpha} -q \phi_c$, of the eigenstates of the
vertical confinement with respect to the conduction band edge in the
centroid layer ($q\varepsilon_{\alpha}$) and of the electrostatic
potential energy ($-q\phi_c$) in the
centroid layer, that is where one can think all charge localized,
following the approach of Refs. \cite{lopezvillanueva,mugnaini}.
For simplicity, $\alpha$ denotes the set of the quantum numbers specifying the
confinement.  
The dimensionality also modifies the Fermi-Dirac integrals $F_{\nu-1/2}$ and $F_v$ entering the
ballistic equations for the mobile charge and the current in the
channel, respectively.
In particular for a 1D conductor, in effective mass
approximation, $v=0$, whereas for a 2D MOSFET
$v=1/2$ \cite{mugnaini2}.
A definition for the Fermi integrals, with $v>-1$, is
\begin{equation}
  F_v(\eta)= \frac{1}{\Gamma(v+1)}\int_0^\infty \frac{x^v}{e^{x-\eta}+1} \ud x 
  \label{eq:FDint}
\end{equation}
with $\Gamma$ being the Gamma function, acting as a normalizing
factor for the Fermi integrals.
For $v \le -1$ we can rely to their property $(\ud/\ud
\eta) F_v(\eta)= F_{v-1}(\eta)$ for their definition \cite{goano}.

We start from a generic multi-subband degenerate version of the ballistic model in
\cite{mugnaini2}.
The vertical electrostatic model we propose, is similar to that in
\cite{jimenez03,jimenez04}, it is also somewhat less
sophisticated, because we will suppose that screening can be
considered constant.
This is done in view of obtaining an analytical
model of intermediate transport.
Anyway we note that this assumption is
sound enough for small cross-section channels and low
electron densities \cite{roldan}.   
Indeed, in the inverse layer centroid approach, we consider the charge
accumulated in the centroid layer and its geometrical screening is included in the effective gate capacitance 
as a series contribution $C_d$, therefore the effective gate oxide capacitance for unit length is given by:
\begin{equation}
  C_g=\left(\frac{1}{C_{ox}}+\frac{1}{C_d}\right)^{-1} 
  \label{cg}
\end{equation}
as reported in Figure \ref{fig:cap}, where the expression $C_{ox}$
depends on the geometry.

If we suppose an undoped channel, consistently with
Fig.\ref{fig:bands}, the linear density of mobile charge 
on the peak of the potential barrier in the channel is given by:
\begin{equation}
  Q_{m}=-C_g\left[ V_g - (\phi_{m}-\chi)/q -\phi_{c}\right],
  \label{eq:qm_el}
\end{equation}
where $\phi_c$ is the electrostatic potential in the centroid layer, 
$\left( \phi_{m}-\chi\right)/q$ is the flat
band potential, given by the difference between the gate workfunction $\phi_m$ and the
channel electron affinity $\chi$.

In the case of ballistic transport, there is no local equilibrium so that no
quasi-Fermi level can be locally defined, because two different carrier
populations exist, originating from source and drain, that can be considered
at equilibrium with the injecting electrodes, as discussed in
\cite{crowell}. 
These two populations are separated by the peak of the barrier in the
channel, that controls transport. 
Therefore, only three points are important: source, drain and the peak
of the electrostatic potential. 
In ballistic transport, carriers move without inelastic scattering along the channel and therefore at the subband
peak the carriers that propagate toward the drain (``forward
states") only come from the source, whereas the carriers propagating toward the
source (``reverse" states) come from the drain. 
As a consequence, we have the superposition of two hemi-Fermi-Dirac distributions.
Following the considerations in \cite{jimenez03}, we can write for the 
ballistic mobile charge linear density on the peak:
\begin{equation}
  Q_m=-q \sum_{\alpha} N_{\alpha} [ F_{-1/2}(\eta_s^\alpha) + F_{-1/2}(\eta_d^\alpha) ],
 \label{eq:qm_inj}
\end{equation}
where: 
$$
N_\alpha= g_\alpha \sqrt{\frac{k_B T m_\alpha}{2\pi^2\hbar^2}}
\Gamma\left(\frac{1}{2}\right) 
$$
is one half of the effective density of states of the
$\alpha$-th subbands multiplied for its degeneration index $g_\alpha$.
$F_{-1/2}(\eta)$ is the Fermi integral of order $-1/2$
and 
\begin{equation}
  \begin{array}{ll}
    \eta_{s}^\alpha=(\phi_c -V_{s} -\varepsilon_\alpha)/\phi_t,\\
    \eta_{d}^\alpha=(\phi_c -V_{d} -\varepsilon_\alpha)/\phi_t.
  \end{array}
\end{equation}
The Fermi potential at the source (drain) is $V_{s(d)}$ and $\phi_t=
K_b T/q$ is the thermal potential.
Equations (\ref{eq:qm_el}) and (\ref{eq:qm_inj}) have to be
solved simultaneously to obtain $\phi_c$ and $Q_m$.

From the Landauer formula, we can write the
current as \cite{jimenez03}:
\begin{equation}
  I_{ds} =q \sum_\alpha G_\alpha [F_0(\eta_s^\alpha) - F_0(\eta_d^\alpha)],
  \label{eq:I_b}
\end{equation}
where $G_\alpha=g_\alpha \frac{k_b T}{\pi \hbar} \Gamma(1)$ is the
effective injection rate of a 1D channel multiplied by the degeneration $g_\alpha$ of
the $\alpha$-th subband. 
Note that the current is dependent on the channel potential through 
$\eta_{s}$ and $\eta_{d}$.

\section{FROM BALLISTIC TO DRIFT-DIFFUSION TRANSPORT}
\input{figureb.tex}

We follow the approach developed in \cite{mugnaini,mugnaini2} for a 2D
MOSFET for the non-degenerate and degenerate cases.
Here we analyze the case of a silicon nanowire transistor (SNWT) where the different dimensionality leads to different Fermi integrals entering
the current and the charge expressions, and to different electrostatics.
Moreover we considered a multi-subband degenerate model while in
\cite{mugnaini2} only a single subband was presented.
We obtain also a correcting factor for the degenerate case, correcting the result of the
linearization process whenever the low field approximation is
not in fully satisfied.   

We recall that, within the B\"uttiker probes approach, inelastic
scattering is thought as localized in special points, spaced by a
length defined as ``mean-free path'' $\lambda$.
The virtual probes act as localized reservoirs along the channel, in which
carriers are thermalized in equilibrium with their quasi-Fermi
potential $V_k$, while transport from one virtual probe to the
next is considered purely ballistic.
We have a drift-diffusion transistor when the channel length is much
longer than the free mean path, that from our point of view it is
equivalent to have a long enough chain of ballistic transistors, as rigorously shown in \cite{mugnaini}.
On the contrary, when the number of internal contacts is small, transport is far-from-equilibrium, and is fully ballistic in the limit $N=1$.
 We remark that within the B\"uttiker probe approach, transport of hot electrons is
accounted only inside each ballistic channel, whereas a full thermalization
occurs in correspondence of each probe, where electron density is described 
by a single quasi-Fermi level. 
It would also be interesting, but out of the 
scope of the present paper, to couple the transport equation with a
heat diffusion equation, accounting for the energy losses in
the B\"uttiker probes, leading to a non uniform temperature
distribution in the device.

We define $V_k$ as the quasi-Fermi potential of the $k$-th virtual probe, and suppose
that the $k$-th contact is placed at $x_{k}=k\lambda$ with
$k=1,\dots,N$, where the boundaries are fixed as $V_0=V_s=0$~V and $V_N=V_d=V_{ds}$. 
That is equivalent to place $N$ ballistic SNWTs of
channel length $\lambda$ in series, as sketched in Figure \ref{fig:chain}.
Since the current $I_k$ in any $k=1,\dots,N$ FET must be equal
to $I_{ds}$, we have $N$ equations determining the local Fermi levels:
\begin{equation}
  I_k =q \sum_{\alpha} G_\alpha [F_{0}(\eta_{k-1}^\alpha)-F_{0}(\eta_k^\alpha)].
    \label{eq:I_k}
\end{equation}
We note that
$\eta_{k}^\alpha=(\tilde{\phi}_{k}-V_{k}-\varepsilon_\alpha)/\phi_t$ where
$\tilde{\phi}_{k}$ is the electrostatic self consistent potential in the
conduction band peak of channel $k$, between the source contact $k-1$ and the drain contact
$k$.
Introducing the definition $\tilde{V}_{k} \equiv (V_{k-1}+V_k)/2$, i.e. the
mean potential between the two contacts ($k$ and $k-1$) of channel
$k$, and making explicit the Fermi integrals, we can rearrange (\ref{eq:I_k}) as follows
\begin{equation}
  I_k = q \sum_\alpha G_\alpha \int_0^\infty
  \frac{\sinh \left( \frac{V_k-V_{k-1}}{2\phi_t}\right) }{ 
    2\left[ \cosh \left( \frac{x-\tilde{\eta}_{k}^\alpha}{2} \right) \right]^2
    + \left[\cosh\left(\frac{V_k-V_{k-1}}{2\phi_t}\right)-1\right]} \ud x
  \label{eq:I_kch}
\end{equation}
where $\tilde{\eta}_{k}^\alpha=(\tilde{\phi}_{k}-\tilde{V}_{k}-\varepsilon_\alpha)/\phi_t$.

At this point, in order to build an analytical model, we consider a
large number of contacts and, having in mind
that the current is constant along the channel, we extend $V_k$ to a
continuous quasi-Fermi potential $V(x)$ satisfying the conditions:
\begin{align}
  V\left(\frac{x_k+ x_{k-1}}{2}\right) & = \tilde{V}_{k}\\
  \frac{\ud V}{\ud x} \left(\frac{x_k+ x_{k-1}}{2}\right) & =\frac{V_k-V_{k-1}}{\lambda}.
\label{eq:cont_approx}
\end{align}
Under the particular hypothesis that every ballistic SNWT works in the linear region i.e. that:
\begin{equation}
  \lambda \frac{\ud V(x)}{\ud x} = V_{k}-V_{k-1}<<2\phi_{t},
  \label{eq:hyplow}
\end{equation}
and expanding the terms $\sinh$ and $\cosh$ to the first order, we can
put (\ref{eq:I_kch}) in the local form
\begin{gather}
  I = \frac{q \lambda}{\phi_t} \frac{\ud V(x)}{\ud x} \sum_{\alpha} G_\alpha F_{-1}[\eta^\alpha(x)]   
\label{eq:I_cont}
\end{gather}
where the quantities pertain to the point $x$, as 
\begin{equation}
  \eta^\alpha(x)=[\phi(x)-V(x)-\varepsilon_\alpha]/\phi_t.
  \label{eq:eta_x}
\end{equation}
Evidently, if a voltage $V_{ds}$ is applied to the chain, (\ref{eq:hyplow})
is satisfied if $V_{ds}<<2\phi_{t}N$. 
Moreover, within the same approximations, the vertical electrostatics becomes:
\begin{equation}
  Q(x) = \sum_\alpha Q^\alpha(x) =  2 q \sum_\alpha N_\alpha  F_{-1/2}[\eta^\alpha(x)].
  \label{eq:Q_loc}
\end{equation}

An important aspect of (\ref{eq:I_cont}) and
(\ref{eq:Q_loc}) is that the current $I^\alpha$ in subband $\alpha$
can be written in terms of the mobile charge density $Q^\alpha(x)$ and of a mobility $\mu^\alpha_{deg}(x)$ as:  
\begin{equation}
  I^\alpha(x) = \mu^\alpha_{deg}(x) Q^\alpha(x) \frac{dV(x)}{dx}
\end{equation}
where the conduction is affected by the 1D electron gas degeneracy
through the mobility $\mu_{deg}(x)$.
The mobility is given by
\begin{equation}
  \mu_{deg}^\alpha(x) =\frac{v_{\alpha}\lambda}{2\phi_t} 
  \frac{F_{-1}[\eta^\alpha(x)]}{F_{-1/2}[\eta^\alpha(x)]}
  \label{eq:mob_deg}
\end{equation}
 where ${v_{\alpha}\lambda}/{2\phi_t}$, to which the expression
(\ref{eq:mob_deg}) is reduced in the non-degenerate limit, represents the low-field
mobility for a 1D gas of incoming electrons described by an hemi Maxwell-Boltzmann
statistics \cite{mugnaini2} occupying the $\alpha$-th subband, whose
mean electron velocity is $v_{\alpha}=\sqrt{\frac{2kT}{\pi m_\alpha}}$, characterized by a ballistic motion for paths of length
$l<\lambda$ and a sudden and complete scattering at $l=\lambda$.
Considering the mean free path $\lambda$ as a constant,
(\ref{eq:mob_deg}) describes the degradation of carrier mobility due to degenerate conditions
of Fermi-Dirac statistics.
An analogous expression was recognized in a Monte Carlo simulation
\cite{watling} for the case of strained silicon FETs.
While, in \cite{wang} and \cite{rahman}, a similar,
but not identical relation between the mean free path and the
effective mobility has been found.

The current is expressed in a local form in (\ref{eq:I_cont}) and we
can eliminate the gradient of the local quasi-Fermi level 
integrating along the channel and exploiting current continuity,
leading to  
\begin{equation}
  I_{ds}= \int_0^L q \frac{\lambda}{\phi_t L} \sum_\alpha G_\alpha
  F_{-1}[\eta^\alpha(x)]  \frac{\ud V(x)}{dx} \ud x. 
    \label{eq:I_int}
\end{equation}
In order to obtain a more compact form of (\ref{eq:I_int}), we can change the integral variable as follows
\begin{equation}
  \int_0^L q F_{-1}[\eta(x)]  \frac{\ud V(x)}{dx} \ud x =
  \int_{V_s}^{V_d} F_{-1}[\eta(V)] \ud V = \int_{\eta_s}^{\eta_d}
      F_{-1}(\eta) \frac{\ud V}{\ud \eta}\ud \eta, 
\end{equation}
where the term $\ud V/\ud \eta$ is obtained by differentiating
(\ref{eq:Q_loc}) in $\ud V$ and using the fact that $\ud \eta/\ud V
=(\ud \phi/\ud V -1)/\phi_t$.
The current can be obtained with a numerical integration of 
\begin{equation}
  I_{ds}^\alpha= \frac{q \lambda G_\alpha}{\phi_tL}  \int_{V_s}^{V_d} F_{-1}[\eta_\alpha(V)] \ud V, 
  \label{eq:Ids_int}
\end{equation}
where we note that $\eta_\alpha$ not only explicitly depends on $V$, but
also implicitly through $\phi_c$, as shown in (\ref{eq:eta_x}).
Such dependence has to be taken into
account in the self-consistent solution of the vertical electrostatics.

Finally, we obtain the compact expression (which we will refer as the DD
model) for the source-drain current for each subband $\alpha$ 
\begin{equation}
  \begin{array}{ll}
    I_{ds}^\alpha = \frac{ q \lambda \phi_t G_\alpha}{\phi_tL} \Big(&
    [F_0(\eta_s^\alpha)-F_0(\eta_d^\alpha)] + \\
    &\sum_\beta \rho_\beta [{\cal F}_{-1,-3/2}(\eta_s^\alpha,\eta_s^\beta)-{\cal F}_{-1,-3/2}(\eta_d^\alpha,\eta_d^\beta)]
    \Big) .
  \end{array} 
  \label{eq:Ids}
\end{equation}
Where for simplicity we defined
\begin{equation}
  {\cal F}_{-1,-3/2}(\eta^\alpha,\eta^\beta)\equiv
  \int_{-\infty}^{\eta^\alpha} F_{-1}(x)
  F_{-3/2}\left(x+\frac{\varepsilon_\beta-\varepsilon_\alpha}{\phi_t}\right) \ud x, 
\end{equation}
with $\rho_\alpha = \frac{q N_\alpha}{C_g \phi_t}$.
It is worth noting that, as observed also in
\cite{mugnaini,mugnaini2}, in the non-degenerate limit the first term in (\ref{eq:Ids})
reduces to the diffusion term of the EKV model, while in the degenerate limit,
it corresponds to the current of a single ballistic channel divided by $N$. 
The second term of (\ref{eq:Ids}) is instead associated with the
drift current.

We note that actually, in the low field approximation, for the case of a
1D channel, the integral (\ref{eq:I_kch}) can be analytically solved
as discussed in Appendix \ref{app2}.
However, the use of the analytical expression leads to the more complex
and numerically expensive expression for the drift-diffusion current (\ref{eq:correction}),
while giving only a slight correction of (\ref{eq:Ids_int}).
The model that employs the analytical solution of (\ref{eq:I_kch}),
with the use of (\ref{eq:correction}), will be referred as DD* model. 
We conclude stressing the fact that DD and DD* models, considered alone, are not appropriate to
describe transport whenever the condition (\ref{eq:hyplow}) is not
satisfied: for example in very short channels, where intermediate transport
is expected.

\section{COMPACT MODEL FOR INTERMEDIATE TRANSPORT}
\input{figurec.tex}

Now we are interested in the
development of a model that will be effective in the whole range of
transport regimes, as proposed in \cite{mugnaini} for the 2D MOSFET case. 
It is evident that in the general case of intermediate transport, the simplifying hypothesis (\ref{eq:hyplow}), 
that enforces each SNWT of the ballistic chain to operate in the
linear region, does not hold, and we can expect that some
elementary channels can work in the saturation regime, or near it \cite{mugnaini}.
The behavior of a SNWT operating in such intermediate transport
regime, can be obtained by solving the system for the complete ballistic chain
(\ref{eq:I_k}), but it can
represent a heavy computational burden, especially for a large number of internal nodes.
In order to build a simple model, that can be 
more easily handled, we note in Fig.~\ref{fig:FL}, where we plotted
the quasi Fermi potential on the virtual probes for a chain decomposition
of a SNWT, that when the saturating behavior of the
elementary ballistic transistor emerges, it is present mainly on the
last ballistic transistor of the chain.
This non-linear behavior is a general condition for transistors in
intermediate transport regime, due to the fact that in its end the
channel narrows down and therefore, to maintain constant the
current flux along it, a major spacing between the last source and
drain levels is required (or, in the continuous limit, a steep drop of the quasi Fermi level). 
This fact suggests that we can aggregate the first $N-1$ ballistic
transistors in an approximate equivalent drift-diffusion transistor 
with ratio $L/\lambda=N-1$ working in low field conditions, as it is represented in
Fig.~\ref{fig:chain}, and similarly to what was proposed in \cite{mugnaini}
for MOSFETs.
We can therefore see a SNWT in intermediate regime as the series of a
drift-diffusion (DD) channel for the first $N-1$ transistor, that we
can solve using (\ref{eq:Ids_int}), and a single ballistic
transistor, governed by (\ref{eq:I_b}).
We only need to solve the DD+B system, imposing constant current
through the two sections in series and solving the Fermi potential on the internal
node between the DD and the B channels.     
We point out that in the DD+B model the ratio $L/\lambda$ has no need to be integer,
because we apply the continuous DD equation (\ref{eq:Ids}).

\section{APPLICATION TO GAA-SILICON NANOWIRE}
\input{figured.tex}

At first we will consider a silicon wire with square cross section of side $t_{si}$.
Concerning the effective gate capacitance, unfortunately no simple
analytical closed form is available.
In order to simplify the electrostatics, we suppose that the interface
between the oxide and silicon is approximately isopotential and
we consider the variational
approximation reported in \cite{liang}, where the capacitance per unit
length of a rectangular coaxial line, associated with the oxide layer, is
\begin{equation}
  C_{ox}=\frac{8 \epsilon_{ox}}{\ln\left(1+\frac{2t_{ox}}{t_{si}}\right)}
\end{equation}
and similarly the capacitance associated with the charge in the
silicon body $C_d$ is
\begin{equation}
  C_{d}=\frac{8\pi\varepsilon_{si}}{\ln\left( \frac{t_{si}}{2z_I}\right) }. 
\end{equation}
The use of $C_d$, with $z_I$ adequately chosen, permits to treat the
capacitance due to the charge distribution in the channel cross
section, in series with the oxide capacitance $C_{ox}$.   
The term $z_I$ is a characteristic dimension of the closed line where
we can effectively localize the whole mobile charge. 
Here it is used as a fitting
parameter for simplicity, while it is actually dependent, due to
volume inversion, on the charge density in the channel.
We point out though that it is smoothly varying for small section
nanowires and low electron density \cite{roldan}.

In case of rectangular quantum confinement, the eigenvalues of the 
Schr\"odinger equation can be considered for simplicity \cite{jimenez03}:
\begin{equation}
  q\varepsilon_{n_x,n_y}^\nu =
  \frac{\hbar^2\pi^2}{2}
  \left[\frac{n_x^2}{m^\nu_x t_{si}^2}+
    \frac{n_y^2}{m^\nu_y t_{si}^2}\right],
\end{equation}
where the mass tensor can be defined as:\\
\begin{center}
  \begin{tabular}{|c||c|c|c|}
    \hline
    $\nu$ & $m_x^\nu$ &  $m_y^\nu$ & $m_z^\nu$ \\ \hline\hline
    $1$   & $m_t$     & $m_l$     & $m_t$ \\ 
    $2$   & $m_l$     & $m_t$     & $m_t$ \\  
    $3$   & $m_t$     & $m_t$     & $m_l$ \\ 
    \hline
  \end{tabular} 
\end{center}
$m_l$ and $m_t$ are the longitudinal and transverse components of
the effective mass tensor of the degenerate minima of the
conduction bands in Si.
We can write the effective mass, characterizing
the motion in the unconfined direction ($z$), as $m_\alpha = m^\nu_z$
for a (100) silicon wire, with $\nu$ running on the different Si
conduction band minima.

In the case of cylindrical quantum confinement, for the gate
capacitance, we have instead:
\begin{equation}
  C_{ox}=\frac{2\pi\varepsilon_{ox}}{\ln\left( 1+\frac{2t_{ox}}{t_{si}}\right) } 
\end{equation}
and
\begin{equation}
  C_{d}=\frac{2\pi\varepsilon_{si}}{\ln\left( \frac{t_{si}}{2z_I}\right) }. 
\end{equation}
Considering cylindrical quantum confinement \cite{jimenez03}, 
we have that the subband separation from the bottom of the conduction
band is described 
by the approximated and handy expression:
\begin{equation}
  q\varepsilon_{n_1,n_2}^\nu \simeq \frac{\hbar^2 \pi^2}{2\sqrt{m_x^\nu m_y^\nu} R^2}
  \left(n_1 + \left| n_2\right| -\frac{1}{4}\right)^2
\end{equation}
where $n_1$ is the radial quantum number, $n_2$ the
azimuthal quantum number and $\nu$ runs on the different silicon valleys.

We applied our model to a cylindrical quantum wire and a
rectangular SNWT, denominated $C$, $R$ respectively. 
Their geometries are described by Fig.~\ref{fig:cap}.
The inversion centroid layer depth was fixed in comparison
with the 2D Schrodinger-Poisson simulator NANOTCAD
\cite{curatola2004}, as shown in Fig.~\ref{fig:VE}.
We note that a careful choice of $z_I$ permits to recover the inverse
layer centroid potential in full agreement with the NANOTCAD
simulator, correctly accounting, thus, for the screening due to the
charge inside the channel as a function of the gate potential.

\section{RESULTS}
\input{figuree.tex}
For a short channel transistor with length of few mean free paths,
transport is quasi ballistic, and we have seen that
the DD model (\ref{eq:Ids_int}) fails to describe its behavior.
On the other hand, it is well known that the transport regime of a transistor with channel length much longer
than the free mean path is described by the drift-diffusion model.
We want to check if our model is able to correctly reproduce such
transition and to investigate the number of free mean paths after
which transport can be definitely associated to the drift-diffusion
regime.
In Fig.\ref{fig:CDD} we plot the output characteristics for a chain of
$N$ ballistic channels numerically calculated (denoted $N$B) and
with the Drift-Diffusion approximation (DD characteristics), for $N=
10$, $20$, $30$ and gate potential $V_g=0.8$ V.
We note that for a SNWT of length smaller than $10 \lambda$, a DD
description is not appropriate.
With increasing $N$ the difference between the $N$B and DD
models is reduced, and for a channel of length $>20\lambda$  the output characteristics calculated with the DD model
fully reproduce the corresponding numerically evaluated $N$Bs.     

We have calculated the output characteristics of SWNTs described in the
latter section employing a direct numerical solution of the chain of
$N$ ballistic transistors ($N$B), and compared them with our models for intermediate transport DD+B and
DD*+B.
Figure \ref{fig:C} and \ref{fig:R1} respectively show the output characteristics for the $C$
SWNT with $N=L/\lambda=5$, and for the rectangular SWNT with $N=2$.
Similar considerations apply to the two figures.
While the DD approximated equation, derived from the linearization of the
$N$B chain, inadequately reproduces the saturation behavior of the $N$B
characteristics, the DD+B model seems suitable to
describe SNWTs in the intermediate transport regime.
As shown in \ref{fig:C} and \ref{fig:R1} the  DD+B and DD*+B models are really able to capture the non-linear behavior of
the $N$B transistors, although a non negligible error remains in the
saturation regime. 
This is due to the the weakly non-linear 
transport in the DD section, that has been neglected.
We note that in general the DD+B* improves the agreement with the
ballistic chain characteristics.

After testing our model with rectangular and cylindrical nanowire
geometries, changing both the oxide and silicon length $t_{ox},
t_{si}$ and  with different values of $N$, we can conclude that the
DD+B and DD*+B compact model quite well reproduces the
output characteristics of degenerate SWNTs for any $N$, with errors in the saturation zone of few percentage points.

In Fig.~\ref{fig:transC} the transfer characteristics of a $C$ SNWT,
treated as a chain of $N$ elementary ballistic channels, with $V_{ds}=0.5$ V, is presented.
Both the DD+B and the DD*+B models well reproduce the behavior of the corresponding
ballistic chain in all gate voltage regimes, for all values of $N$. 
We note that the DD*+B model is always more accurate, in particular the
correction is more evident for transistor with few nodes, at large
gate voltage.

We investigated also the so called ballisticity index of a transistor \cite{natori08}, that
is given by the current ratio $I/I_b$, between the current of the
transistor and that of a corresponding ballistic one.
The results of its calculation for a $N$B chain, with $N$ ranging from
$1$ to $20$, are shown in Fig.\ref{fig:Bindex}, where we considered
the $R$, $C$ SNWTs and also, for comparison an
undoped Double Gate MOSFET with $t_{si}=4$ nm, $t_{ox}=2$ nm.
The ballisticity index is monotonous and slowly decaying with $N$, the behavior is
similar for all the transistors considered here.
The curves can be easily fitted with the function $1/\left[1+r(N-1)\right]$
where $N$ is the number of ballistic elements and $r\approx0.25$.
We note that initially the ballisticity index steeply decreases with
$N$.
For longer channels the current becomes slowly varying with
$N$: sliding from $N=10$ to $N=20$ the current only decreases from the
$30$\% to the $20$\% of the ideal ballistic one.

Having in mind a compact model, the calculation of equations (\ref{eq:Ids}) or (\ref{eq:Ids_int}) for
the DD+B model are still computationally expensive.  
Therefore we also tested the approximation of the integral in the
drift-diffusion section (DD) with its symmetrical linearization
\cite{gildenblat_sym}, as discussed in Appendix \ref{app1}.

\section{CONCLUSION}
We have presented a physics-based analytical model able to describe
quasi one-dimensional field-effect transistors in the complete
range of transport regimes extending from the fully ballistic case
captured by the Natori model to the quasi-equilibrium case captured
by the drift-diffusion description.
Our proposed model sees a generic transistor as a
long enough chain of elementary ballistic transistors in series with a
common gate. Based on the B\"uttiker probes description of inelastic
scattering, we have rigorously proved that the model reduces to the
limit cases.
In addition, as the most important result in this paper, we have shown
that an equally adequate model, much simpler from the computational 
point of view,
and more physically intuitive, is represented by the series of an 
appropriate
drift-diffusion one-dimensional transistor and a ballistic 
one-dimensional transistor,
consistently with the results in \cite{mugnaini, mugnaini2}, that apply 
to 2DEG FETs.
We have focused in this paper on silicon nanowire transistors,
but our model is applicable without significant variations to any
type of quasi-one dimensional FET, such as those based on carbon 
nanotubes, graphene
nanoribbons, or other channel materials.

Finally, we have shown that an interesting consequence of our model is 
that,
if a uniformly spaced chain is assumed, the Fermi-Dirac statistics degrades
the low-field mobility, consistently with the
observations in \cite{watling}.

\appendices


\section{ANALYTICAL SOLUTION OF THE DD INTEGRAL \label{app2}}
We note that in the integral
\begin{equation*}
  \int_0^\infty  \frac{1}{ 
    2\left[ \cosh \left( \frac{x-\tilde{\eta}^\alpha}{2} \right) \right]^2
    + \left[\cosh\left(\frac{V_k-V_{k-1}}{2\phi_t}\right)-1\right]} \ud x,
\end{equation*}
in (\ref{eq:I_kch}) has an analytical solution given by 
\begin{align*}
  {\mathcal I}(\tilde{\eta}) &= \frac{1}{\sqrt{a(a+1)}} \times \\  
  &\left\{\tanh^{-1}\left[\sqrt{\frac{a}{a+1}}\right]+\tanh^{-1}\left[\sqrt{\frac{a}{a+1}}\tanh\left(\frac{\tilde{\eta}^\alpha}{2}\right)\right]\right\},
\end{align*}
where $a=\left[\cosh\left(\frac{V_k-V_{k-1}}{2\phi_t}\right)-1\right]/2$.
We replace the Fermi level difference between neighbor probes, by  its
mean value on the linearized chain 
$$ 
V_k-V_{k-1} \approx \gamma=\Delta V^{(DD)}/(2\phi_tN),
$$
where $\Delta V^{(DD)}$ is the total potential drop in the DD
section and $N$ the number of elementary channels in it.

In the low-field approximation we also replaced the
$\sinh(\frac{V_k-V_{k-1}}{2\phi_t})$ with its arguments: we try to
amend this by including a correction factor obtained by the ratio of the not-approximate term over
approximate one
\begin{equation}
  \sinh\left(\frac{V_k-V_{k-1}}{2\phi_t}\right)/\frac{V_k-V_{k-1}}{2\phi_t}.
  \label{eq:corr2}
\end{equation}
In the end, we reach a more accurate version of (\ref{eq:Ids_int})
for the DD section, given by the following expression 
\begin{equation}
I^\alpha =  \frac{q\lambda G_\alpha}{\phi_t
  L}\frac{\sinh(\gamma)}{\gamma \sqrt{a(a+1)}}
  \int_{V_s}^{V_d} {\mathcal I}[V] \ud V
\label{eq:correction}
\end{equation}

\section{SYMMETRICAL LINEARIZATION OF THE DD INTEGRAL IN SNWT \label{app1}}
\input{figuref.tex}

The integral for the drift-diffusion current (\ref{eq:Ids_int}) is
computationally expensive for a compact model to be included in
circuit simulators such as SPICE.
Therefore we adopt a variant of the symmetrical linearization
\cite{gildenblat_sym,gildenblat_deg} in order to obtain an
approximated result:
\begin{align*}
  I & = \int_{V_{s}}^{V_{d}} q\sum_\alpha G_\alpha  
  F_{-1}(\frac{\phi_c-V-\varepsilon_\alpha}{\phi_t})
  \frac{\ud V}{\ud x} \frac{\lambda}{L} \ud x= \\
  & = \int_{\phi_{cs}}^{\phi_{cd}} q\sum_\alpha G_\alpha 
  F_{-1}(\frac{\phi_c-V-\varepsilon_\alpha}{\phi_t}) 
  \frac{\ud V}{\ud \phi_c} \frac{\lambda}{L} \ud \phi_c= \\
  & \simeq q \sum_\alpha G_\alpha 
  F_{-1}(\frac{\phi_{c,m}-V-\varepsilon_\alpha}{\phi_t}) 
  \left(\frac{\ud V}{\ud \phi_c}\right)_m \frac{\lambda}{L} [\phi_{cd}-\phi_{cs}] = \\
  & \simeq q \frac{\lambda}{L} \sum_\alpha G_\alpha
  F_{-1}( \frac{\phi_{c,m}-V-\varepsilon_\alpha}{\phi_t}) 
  [\phi_{cd} - \phi_{cs}] n_q
  \label{IDS_symlin}
\end{align*}
where we have defined the ``quantum slope factor'':
\begin{equation}
  n_{q} \equiv 1+\frac{1} {q\sum_\alpha \rho_{\alpha}
    F_{-\frac{3}{2}}\left(\frac{\phi_{c,m} - V_{c,m} - \varepsilon_\alpha}{\phi_{t}}\right) }
\end{equation}
that is a constant in the considered case.
The linearization is done around:
\begin{equation}
  \phi_{c,m}=\frac{\phi_{cs}+\phi_{cd}}{2}
\end{equation}
where $\phi_{cs}$ and $\phi_{cd}$ can be obtained solving the vertical
electrostatics (\ref{eq:Q_loc}).
Moreover from vertical electrostatics we find $V_{c,m}$ with an iterative process.
We can observe in Fig.\ref{fig:Lin} that the symmetrical
linearization of the DD integral well reproduce the not-approximated
results.

\bibliographystyle{IEEEtran}
\bibliography{IEEEabrv,bibf}


\end{document}

%% file: figurea.tex
\begin{figure}
  \includegraphics[width=6cm]{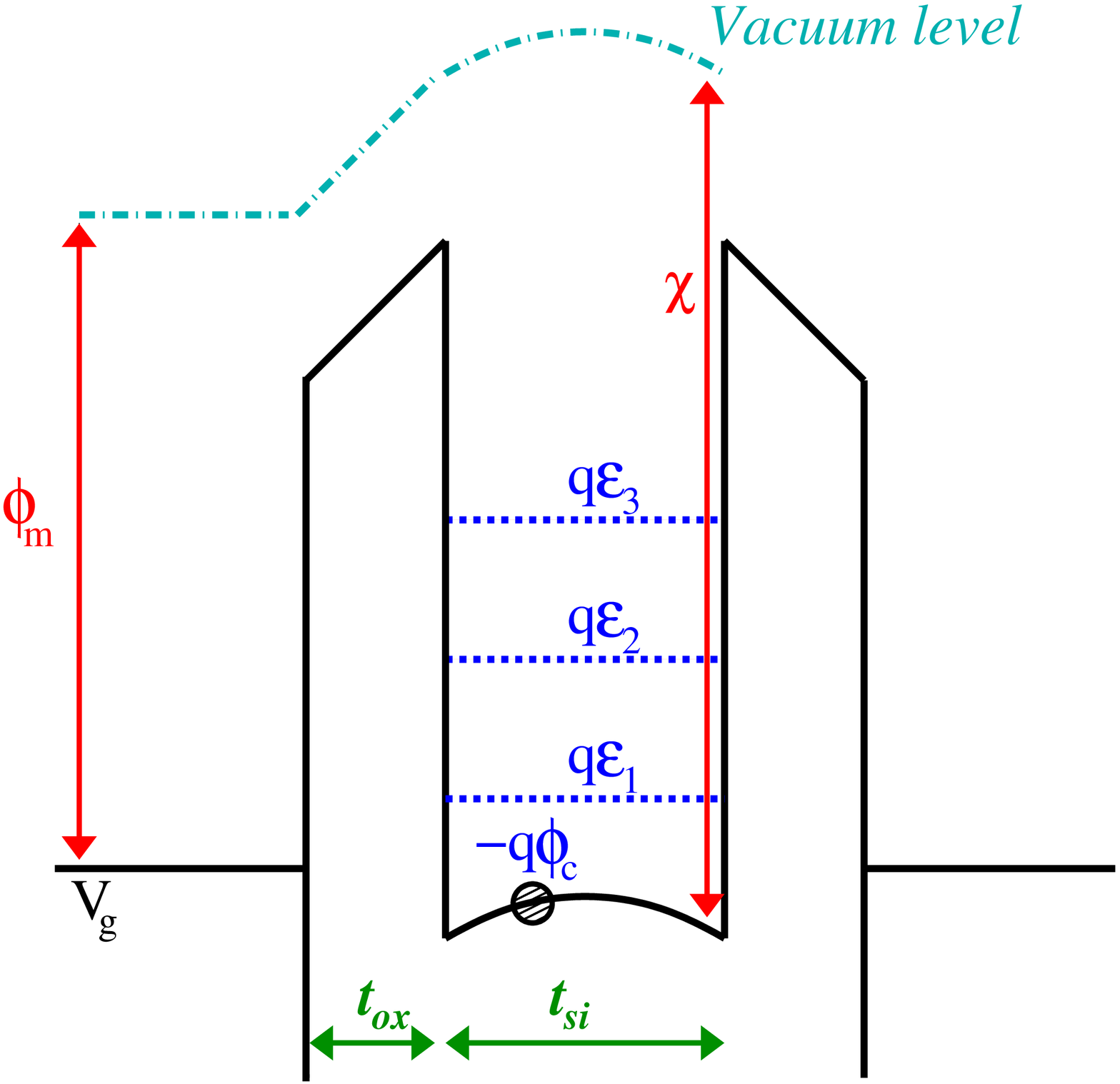}
  \caption{Schematic band diagram of a 1D SNWT. 
    $V_g$ is the gate Fermi
    potential, $\phi_m$ is the gate workfunction, $\chi$ is the channel electron affinity,
    $\phi_c$ is the potential in the centroid layer.
    $q\varepsilon_\alpha$ are the eigenvalues of the vertical
    confinement, drawn out of scale.}\label{fig:bands}
\end{figure}

\begin{figure}
  \includegraphics[width=8cm]{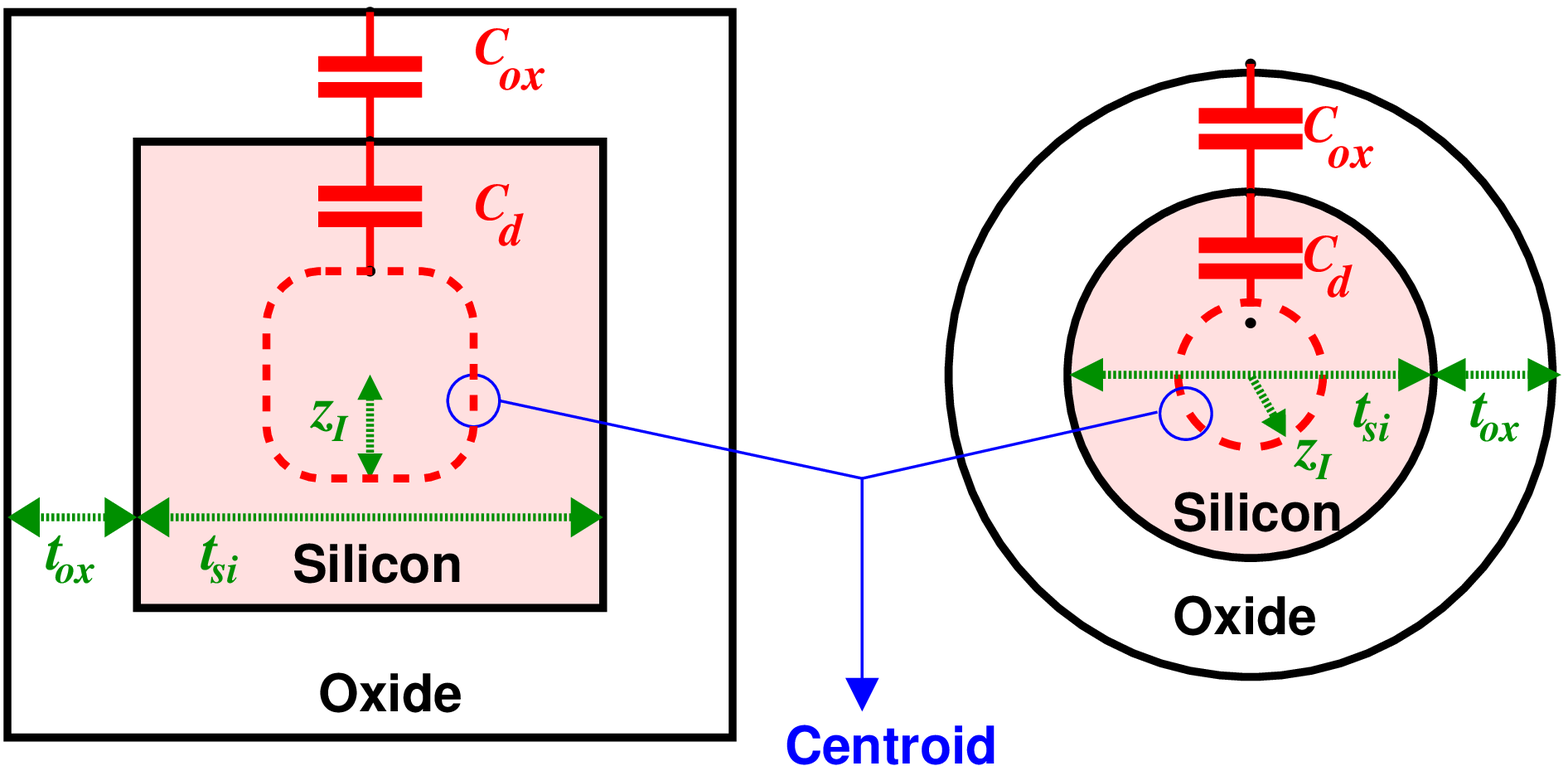}
  \begin{center}
    \vspace{0.5cm}
    \begin{tabular}{|c||c|c|c|}
      \hline
      &  $t_{si}$ &  $t_{ox}$  &  $z_I$    \\ \hline\hline
      $R$  &  $4$ nm   &  $3$ nm     &  $0.97$ nm \\ 
      $C$   &  $4$ nm   &  $3$ nm     &  $0.92$ nm \\  
      \hline
    \end{tabular}
  \end{center}
  \caption{Schematic capacitance diagram for a wire with square and
    circular cross section, representing the series of the oxide
    capacitance $C_{ox}$ and the silicon capacitance $C_d$.
    In the square shape the interface between the silicon and the
    insulator is considered approximately isopotential and the mobile
    charge layer (the centroid $z_I$) is approximated with a square contour.
    The table summarizes the geometrical parameters used in the
    simulation of the rectangular wire ($R$) and cylindrical one ($C$).}\label{fig:cap}
\end{figure}

%% file: figureb.tex
\begin{figure}
  \includegraphics[width=7cm]{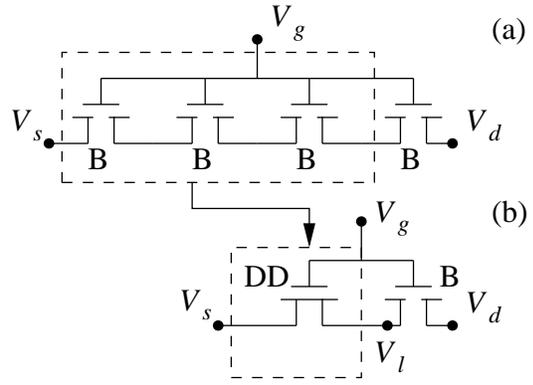}
  \caption{(a) Circuit model of a generic SNWT, subject to inelastic
    scattering, in terms of a convenient chain of ballistic (B) SNWTs.
    (b) Approximate aggregation of the first $N-1$ ballistic
    transistors in an equivalent Drift-Diffusion (DD) one. 
    The macromodel DD+B comes out to be a suitable model for a device in
    intermediate transport regime.}\label{fig:chain}
\end{figure}

%% file: figurec.tex
\begin{figure}
  \includegraphics[width=8.5cm]{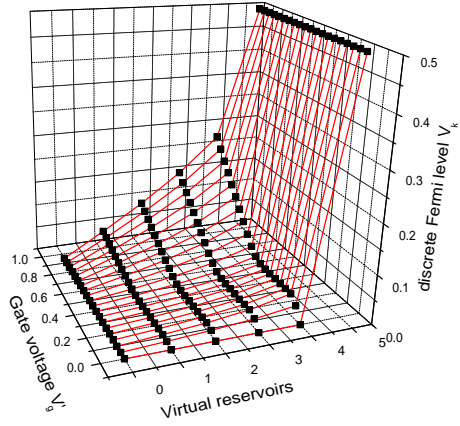}
  \caption{Discrete quasi-Fermi potential for a chain of $N=5$ SNWTs at
    fixed $V_{ds}=0.5$ V. 
    Fermi level is defined only on the virtual
    probes placed at points $x= k\lambda$ with $k=1\dots N$.
    It is evident that the first $N-1$ transistors work
    approximately in linear regime, while increasing $V_g$ a
    non-linear behavior is developed on the latter ballistic channel.
    This fact supports the choice of the DD+B segmentation in our model for
    intermediate transport.}
  \label{fig:FL}
\end{figure}   

\begin{figure} 
  \includegraphics[width=8.8cm]{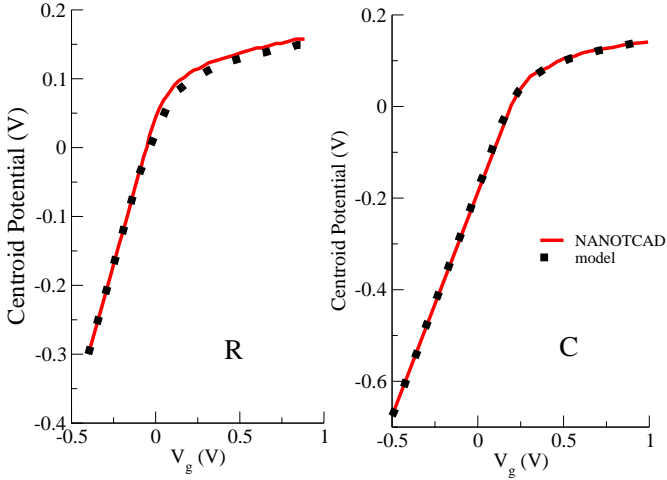}
  \caption{Comparison between NANOTCAD 2D simulation and the compact model
    calculation of the vertical electrostatic for the $R$ and $C$ SNWTs.
    The centroid potential obtained with NANOTCAD 2D (solid line)
    and the centroid potential obtained with the compact model
    (dotted) are reported.
    The compact model well accounts for the centroid
    potential, point in which the charge of the channel can thought to
    be concentrated.}
  \label{fig:VE}
\end{figure}

%% file: figured.tex
\begin{figure} 
  \includegraphics[width=8.5cm]{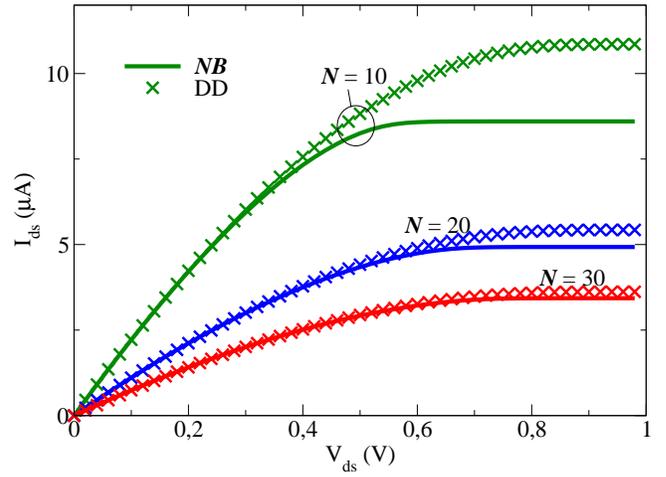}
  \caption{Output characteristics for a chain of $N$ ballistic
    cylindrical $C$ SNWTs (denoted by $N$B), for $N= 10$, $20$, $30$,
    compared withe results of the DD model  with $\lambda=L/N$.
    The gate potential value is $V_g=0.8$ V.
    As the number of ballistic transistors in series increases the
    drift-diffusion approximation become more and more able to capture
    the behavior of the device.}
  \label{fig:CDD}
\end{figure}

%% file: figuree.tex
\begin{figure}   
  \includegraphics[width=8.5cm]{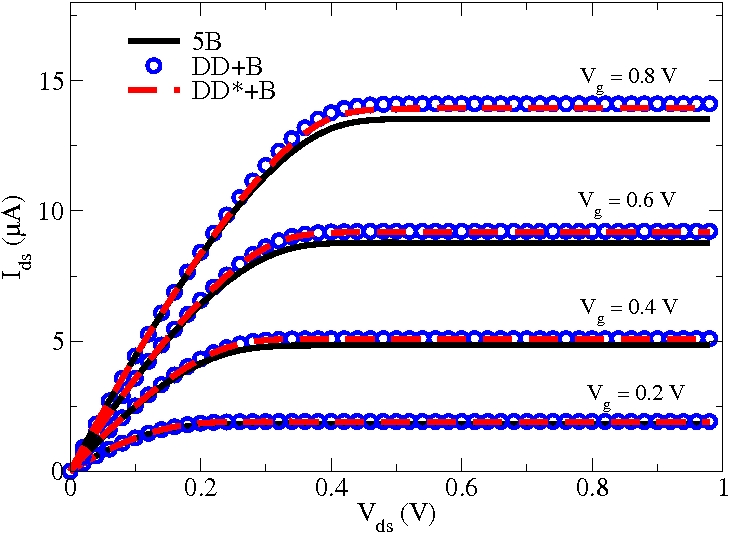}
  \caption{Output characteristics of a C SNWT
    modeled as a chain of $N=5$ elementary transistors: $5$B
    corresponds to the the exact numerical evaluation.
    The DD+B compact model is obtained considering the series of a
    DD channel governed by (\ref{eq:Ids}) plus a ballistic one and the
    DD*+B is analogous but uses (\ref{eq:correction}) for the DD section.
    The choices of the gate potential are $V_g=0.2$, $0.4$, $0.6$,
    $0.8$ V.}
  \label{fig:C}
\end{figure}    

\begin{figure}   
  \includegraphics[width=8.5cm]{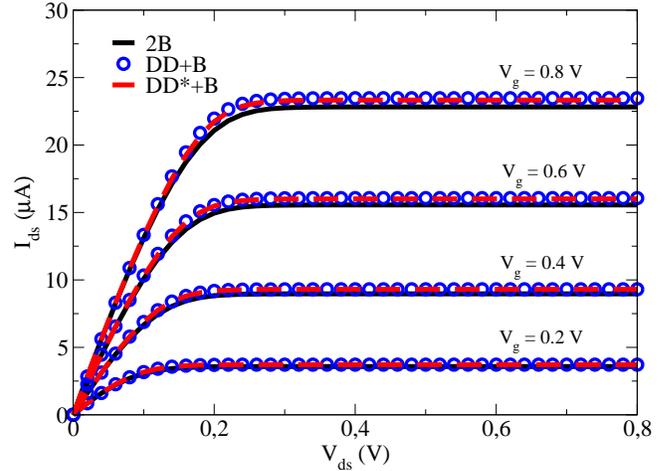}
  \caption{Output characteristics of the R SNWT
    modeled as a chain of $N=2$ elementary transistors:  exact
    numerical evaluation ($2$B), compact model DD+B with a drift-diffusion plus a
    ballistic channels and DD*+B analogous to the latter except for
    the use of (\ref{eq:correction}) in the DD section.
    The choices of the gate potential are $V_g=0.2$, $0.4$, $0.6$,
    $0.8$ V.}
  \label{fig:R1}
\end{figure}     

\begin{figure}   
  \includegraphics[width=8.5cm]{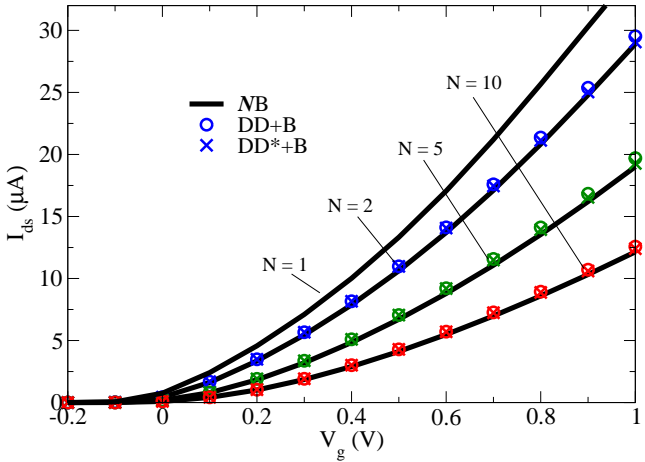}
  \caption{Transcharacteristic curves for a chain of $N$ ballistic C
    SNWTs, with $N$ ranging from $1$ to $10$ and $V_{ds}=0.5$ V.
    The exact numerical evaluation ($N$B), and the results of the DD+B and DD*+B
    compact models are shown.}
  \label{fig:transC}
\end{figure}   
  
\begin{figure}   
  \includegraphics[width=8.5cm]{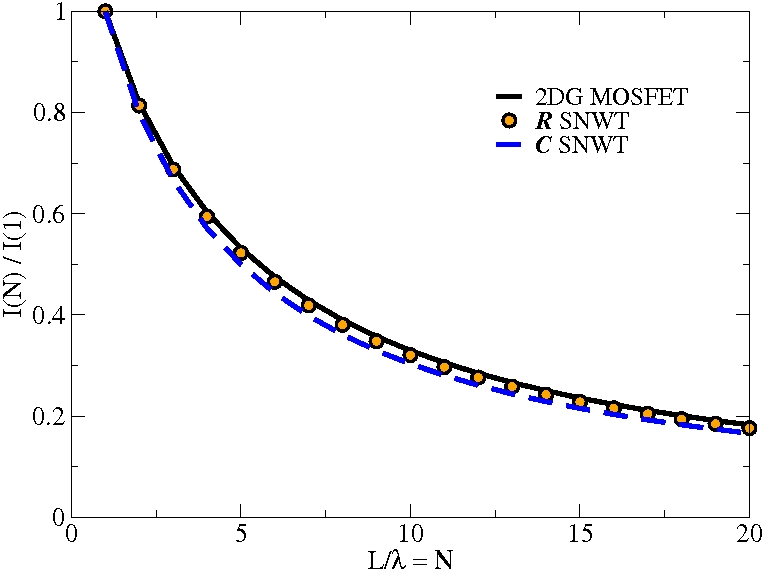}
  \caption{Ballisticity index of a $N$B chain as a function of $N$ for
    the $R$, $C$ SNWTs, and a MOSFET (see text for details).
    The gate potential is $V_g=0.8$V and the drain-source potential is
    $V_{ds}=0.5$V.}
  \label{fig:Bindex}
\end{figure}    

%% file: figuref.tex
\begin{figure}   
  \includegraphics[width=8.5cm]{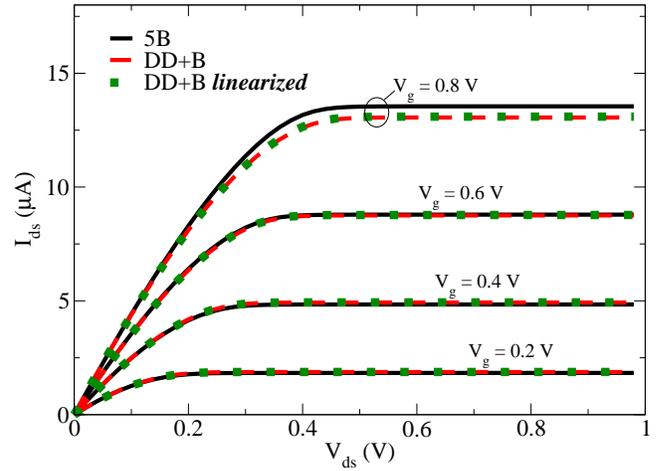}
  \caption{Characteristic curves calculated for a $R$ SNWT with $N=5$.
    The results of the direct numerical calculation $5$B, of the DD+B
    compact model and of the same compact model, employing symmetrical
    linearization DD$_{lin}$+B, discussed in the text, are shown.
    The curves are calculated for gate potential values of $V_g=0.2$,
    $0.4$, $0.6$, $0.8$ V.}
  \label{fig:Lin}
\end{figure}